\documentclass[sigconf]{acmart} 
\definecolor{brown}{rgb}{0.59, 0.29, 0.0}
\definecolor{darkgray}{rgb}{0.59, 0.59, 0.59}
\definecolor{tablegray}{gray}{.9}

\newcommand\rv[1]{\textcolor{black}{#1}}

\newcommand{\system}{\textsc{ContextXR}}
\newcommand{\problem}{context-aware functionality suggestion in XR}
\newcommand{\cost}{\rv{navigation \& search cost}}

\usepackage[utf8]{inputenc}
\usepackage{diagbox}
\usepackage{colortbl}

\newsavebox{\promptboxbox}

\newenvironment{promptbox}{%
  \par\addvspace{1.2em}
  \par\small\noindent
  \begingroup
  \setlength{\fboxsep}{5pt}%
  \setlength{\parindent}{0pt}%
  \setlength{\parskip}{0.25\baselineskip}%
  \begin{lrbox}{\promptboxbox}%
  \begin{minipage}{\dimexpr\linewidth-2\fboxsep-2\fboxrule\relax}
  \ttfamily\small\raggedright
}{%
  \end{minipage}%
  \end{lrbox}%
  \fcolorbox{gray!60}{gray!5}{\usebox{\promptboxbox}}%
  \endgroup
  \par\addvspace{1.2em} % space after each promptbox
}

\usepackage{color}
\usepackage{soul}
\usepackage{fontawesome5}
\usepackage{xcolor}
\usepackage{makecell}
\newcommand{\costdrop}[1]{\textcolor{green!50!black}{#1}}

\usepackage{tabularx}

\usepackage{nameref}
\usepackage{booktabs}
\usepackage{multirow}

\usepackage{xspace}
\usepackage{enumitem}
\usepackage{mathtools}
\usepackage{commath}
\usepackage{algorithm}
\usepackage{algpseudocode}

\usepackage{color-edits}

\usepackage{tikz}
\usetikzlibrary{arrows.meta, positioning}

\usepackage{amssymb}% http://ctan.org/pkg/amssymb
\usepackage{pifont}% http://ctan.org/pkg/pifont
\usepackage{amsmath}

\newcommand{\customtilde}{{\raise.17ex\hbox{$\scriptstyle\sim$}}}

\newcommand{\etal}{et~al.\xspace}
\newcommand{\eg}{e.\,g.,\xspace}
\newcommand{\ie}{i.\,e.,\xspace}

\usepackage{xparse}
\sethlcolor{yellow}

\AtBeginDocument{%
  }

\copyrightyear{2026}
\acmYear{2026}
\setcopyright{cc}
\setcctype{by}
\acmConference[UIST '26]{The 39th Annual ACM Symposium on User Interface Software and Technology}{November 02--05, 2026}{Detroit, MI, USA}
\acmBooktitle{The 39th Annual ACM Symposium on User Interface Software and Technology (UIST '26), November 02--05, 2026, Detroit, MI, USA}
\acmDOI{10.1145/3830398.3830675}
\acmISBN{979-8-4007-2856-3/2026/11}

\begin{document}

% \title{Benchmark-based Evaluation of XR User Interfaces}
\title{A Benchmarking Framework for Context-aware XR Interfaces}

\author{Hyunsung Cho}
% \email{hyunsung@cs.cmu.edu}
\orcid{0000-0002-4521-2766}
\affiliation{%
  \institution{Carnegie Mellon University}
  \city{Pittsburgh}
  \state{Pennsylvania}
  \country{USA}
}

\author{Sarah Yewon Yun}
% \email{ysyun@andrew.cmu.edu}
\orcid{0009-0000-7471-7061}
\affiliation{%
  \institution{Carnegie Mellon University}
  \city{Pittsburgh}
  \state{Pennsylvania}
  \country{USA}
}

\author{Nancy Ruonan Sun}
% \email{ruonans@andrew.cmu.edu}
\orcid{0009-0007-9677-6055}
\affiliation{%
  \institution{Carnegie Mellon University}
  \city{Pittsburgh}
  \state{Pennsylvania}
  \country{USA}
}

\author{Ben Lafreniere}
\orcid{0000-0002-0546-0466}
\affiliation{%
  \institution{Reality Labs Research, Meta}
  \city{Toronto}
  \state{Ontario}
  \country{Canada}
}
% \email{ben.lafreniere@gmail.com}

\author{Mark Parent}
% \email{}
\orcid{0009-0004-2361-3245}
\affiliation{%
  \institution{Reality Labs Research, Meta}
  \city{Toronto}
  \state{Ontario}
  \country{Canada}
}

\author{Kashyap Todi}
% \email{}
\orcid{0000-0002-6174-2089}
\affiliation{%
  \institution{Reality Labs Research, Meta}
  \city{Redmond}
  \state{Washington}
  \country{United States}
}

\author{Tanya R. Jonker}
\orcid{0000-0001-8646-5076}
\affiliation{%
  \institution{Reality Labs Research, Meta}
  \city{Redmond}
  \state{Washington}
  \country{United States}
}
% \email{tanyarjonker@gmail.com}

\author{Hrvoje Benko}
% \email{benko@meta.com}
\orcid{0000-0002-2059-3558}
\affiliation{%
  \institution{Reality Labs Research, Meta}
  \city{Redmond}
  \state{Washington}
  \country{United States}
}

\author{Tongshuang Wu}
% \email{sherryw@cs.cmu.edu}
\orcid{0000-0003-1630-0588}
\affiliation{%
  \institution{Carnegie Mellon University}
  \city{Pittsburgh}
  \state{Pennsylvania}
  \country{USA}
}

\author{David Lindlbauer}
% \email{davidlindlbauer@cs.cmu.edu}
\orcid{0000-0002-0809-9696}
\affiliation{%
  \institution{Carnegie Mellon University}
  \city{Pittsburgh}
  \state{Pennsylvania}
  \country{USA}
}

\renewcommand{\shortauthors}{Cho et al.}


\begin{abstract}
Everyday Extended Reality (XR) systems aim to provide context-aware access to the right functionalities at the right time and place, with minimal manual reconfiguration as users switch context. 
\rv{Yet these interfaces are hard to evaluate: current prototyping and user-study workflows offer no systematic, repeatable way to compare adaptation methods across users and scenarios.}
We present \rv{\system{},} a novel benchmarking framework for context-aware XR interfaces.
\rv{\system{} represents an XR application as a connected graph of \emph{functional facets}, each a semantically coherent group of related capabilities that together support a shared user intent.}
\rv{On this representation, we build} MineXR++, a dataset augmenting prior XR interface data with facet-level annotations, \rv{and formulate three canonical tasks of context-aware suggestion: context factor analysis, initial facet suggestion, and next facet suggestion.}
\rv{Our evaluation protocol scores suggestion methods by a simulated interaction metric, the navigation and search cost of reaching the desired functionality.
Through experiments benchmarking global popularity, relational retrieval, and LLM-based methods, we demonstrate that \system{} enables the systematic, reproducible evaluation of context-aware XR interfaces.}
\end{abstract}



\begin{teaserfigure}
  % \vspace{-0.5em}
  \centering
  \includegraphics[width=0.97\textwidth]{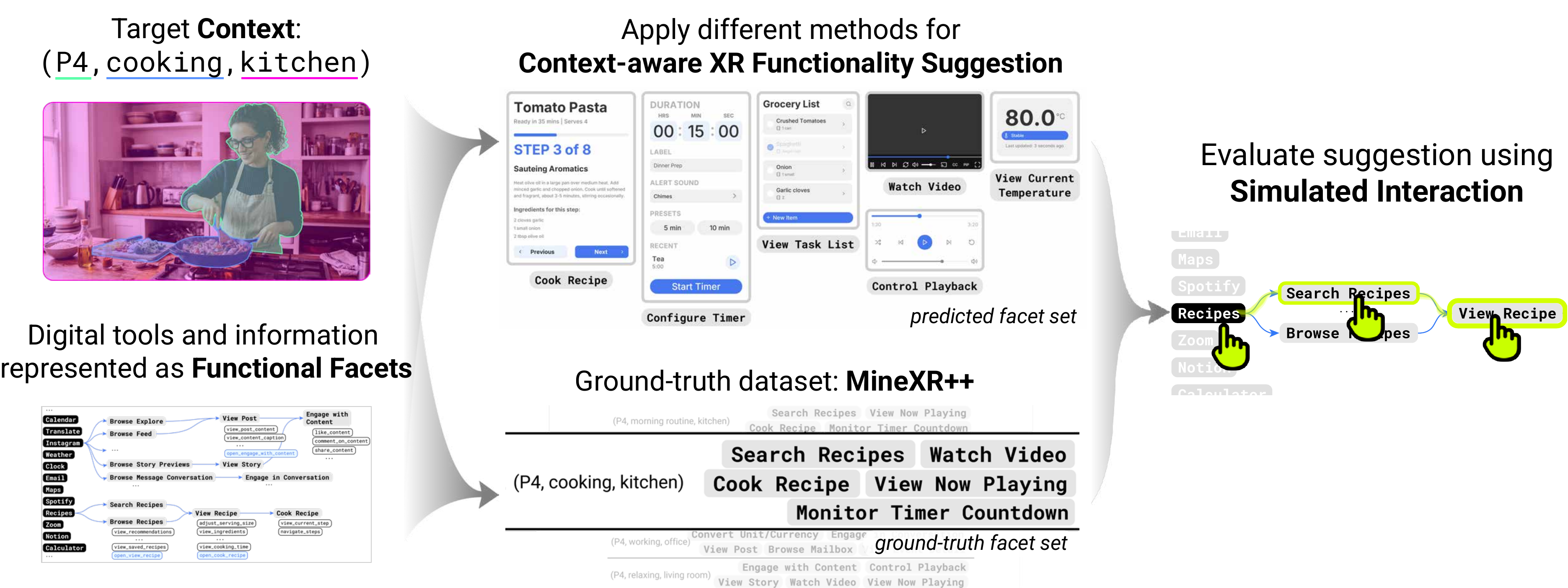}
  \vspace{-0.5em}
  \caption{We propose \system{}, a benchmarking framework for developing and evaluating context-aware functionality suggestion \rv{in XR}. 
  We contribute (1) a representation of digital tools and information as \textbf{functional facets}, (2) \textsc{MineXR++}, an augmented dataset that serves as ground truth, (3) a problem formulation \rv{of the} canonical tasks of context-aware suggestion, and (4) an evaluation protocol \rv{built on a simulated interaction metric}. Jointly, our approach enables principled evaluation of context-aware XR systems.} 
  % \todo{Figure: Evaluate suggestion using simulated interaction metrics}
  \Description{}
  \label{fig:teaser}
\end{teaserfigure}

\maketitle

\section{Introduction}
Extended Reality (XR) \rv{is emerging as an everyday companion.
Worn continuously through glasses and other head-worn displays, XR can offer situated support, facilitating access to digital content right where a task is happening. 
While cooking, for instance, the current step of a recipe and a countdown timer can rest on the kitchen counter, in view and within reach as the user works.}

\rv{A user's context shifts constantly across a day, from cooking to a commute to focused work, and so does the XR functionality that is relevant at each moment. 
Without a system that adapts automatically, users must rearrange the interface by hand, dismissing functionality from the previous context and then searching through apps to surface what the new context needs.
A large body of HCI research has studied how interfaces should adapt to the user, task, and environment to reduce this manual overhead~\cite{grubert2017towards,todi2023framework,lindlbauer2022future}.
The goal is for XR to be \emph{context-aware}, using context to surface content relevant to what the user is doing~\cite{dey1999towards}.}

\rv{A core part of context awareness is deciding \emph{what} functionality to surface, and at \emph{what granularity}. 
Surfacing whole applications exposes more than a situated task requires and quickly clutters the display space~\cite{cho2024minexr,han2023blendmr}.
The right unit is instead a working set of \emph{XR functionalities} for the activity at hand---while cooking, the current recipe step, a timer configuration, and a music control, kept available for as long as the task lasts~(\autoref{fig:teaser}).}
We frame this as \textbf{context-aware functional suggestion in XR}: given a user context, an XR system should infer which functional units ought to be made readily accessible so that desired functionality can be reached with \rv{little navigation and search effort}.

Despite growing interest in context-aware XR, the field lacks the infrastructure to systematically evaluate and compare solutions. 
Existing \rv{systems are typically bespoke prototypes} evaluated through one-off \rv{user studies}~(\eg~\cite{lindlbauer2019context, cheng2021semanticadapt, li2024situationadapt}). 
These are essential for demonstrating end-user value, but \rv{cannot compare} methods across users, scenarios, and modeling strategies in a systematic way, \rv{nor isolate} which contextual factors actually \rv{matter}. 
\rv{Other} fields \rv{have addressed this with shared} evaluation frameworks, \rv{from} computational UI design ~\cite{deka2017rico,wang2021screen2words} to computer vision~\cite{russakovsky2015imagenet,lin2014microsoft} and machine learning~\cite{wang2018glue,wang2019superglue}.
\rv{Such frameworks complement} user studies, \rv{letting} developers \rv{build and compare} solutions against standardized baselines without a costly user study \rv{at every iteration}.
What is missing in context-aware XR is such a benchmarking framework: a shared \rv{representation of} interface functionality, benchmark tasks reflecting realistic support settings, and evaluation metrics that capture user-facing utility rather than prediction \rv{accuracy} alone.

In this paper, we present \textbf{\system{}}, a benchmarking framework for studying context-aware functional suggestion in XR. 
\system{} represents application functionality as \emph{functional facets}, semantically coherent \rv{
groupings of atomic capabilities that support a shared user intent.
These are coarser than a single capability but finer than a whole application.}
% groupings of capabilities that support a shared user intent. 
% These} are \rv{finer} than whole applications but \rv{coarser} than atomic capabilities\rv{, which are too fine-grained to be useful on their own}.
For example, a music app might \rv{expose a} \texttt{Control Playback} \rv{facet}, composed of atomic capabilities such as \texttt{play}, \texttt{pause}, \texttt{stop}, or \texttt{next song}.
A recipe app might \rv{expose a} \texttt{View Recipes} \rv{facet}, composed of atomic capabilities such as \texttt{view\_ingredients}, \texttt{view\_preparation\_steps}, and \texttt{view\_cooking\_time}.
Functional facets \rv{thus} capture what an \rv{end} user is trying to accomplish while remaining grounded in executable interface behavior.
Facets \rv{with}in an application are typically connected (\eg \texttt{Browse Recipes} $\rightarrow$ \texttt{View Recipe} $\rightarrow$ \texttt{Cook Recipe}), which we model as an application's \emph{facet graph}.
Functional facets and the facet graph serve as the computational units of context-aware XR support for retrieval, suggestion, and evaluation. 

\rv{We instantiate functional facets in} \textsc{MineXR++}, an augmented version of the open-source MineXR dataset~\cite{cho2024minexr} annotated with facet-level structure.
\rv{Over this dataset, we define} a set of benchmark tasks for \problem{}: context factor analysis, initial facet suggestion when a user first enters a context, and next facet suggestion during ongoing interaction. 
Finally, we introduce a simulated interaction protocol centered on \emph{\rv{\cost{}}}\rv{, which} quantifies the effort a user \rv{spends} reaching desired functionality given the surfaced suggestions. 
This \rv{protocol} moves beyond exact-match retrieval by accounting for partial utility, such as when a suggested facet provides a useful entry point within the correct application.

We demonstrate \rv{the framework through} benchmark experiments \rv{comparing representative approaches on these tasks.}
These experiments show that \rv{\system{} makes} context-aware XR support computationally tractable and systematically comparable, enabling principled study of which contextual factors matter, how different methods \rv{compare on the same task}, and how much they reduce end-user effort relative to unassisted use.

In summary, we contribute \system{}, a benchmarking framework for studying \problem{}, combining (1) a new representation of application functionality based on \emph{functional facets}, (2) \textsc{MineXR++}, an augmented dataset with facet-level annotations, (3) benchmark task formulations for \rv{context factor analysis,} initial \rv{facet suggestion} and next facet suggestion, and (4) an evaluation protocol centered on \rv{\cost{}}. 
\rv{Our code, dataset, and benchmark are available at: \textit{\href{https://augmented-perception.org/publications/2026-contextXR.html}{augmented-perception.org/publications/2026-contextXR}}.}
\section{Related Work}
\label{sec:related}
We situate our work at the intersection of context-aware suggestions in XR, computational approaches for structured UI development, and adaptive XR interfaces.
These areas motivate the need for finer-grained representations and the value of structured evaluation for context-aware interface support.

\subsection{Context-aware Suggestions in XR}
\rv{Following Dey and Abowd~\cite{dey1999towards}, we treat a system as context-aware if it uses context to provide information or services relevant to the user's task.
\system{} studies one such behavior: using context to suggest relevant XR functionality.}
Prior work on context-aware suggestions in XR has largely focused on proactive assistance at the level of whole applications, single follow-up actions, or specialized task support.
At the application level, COBO~\cite{yu2022optimizing} studies when suggestions should be presented, Lu \etal~\cite{lu2024did} examine how reopening a contextually relevant subset of apps supports users across spatial transitions, \rv{and Langerak \etal~\cite{langerak2026xaiui} generate tailored explanations for such suggestions.} 
However, application-level access is often too coarse for situated use: users typically need only a task-relevant subset of functionality, placed in context, rather than unrestricted access to entire applications~\cite{cho2024minexr,han2023blendmr}.

Other work has explored leveraging contextual information to proactively suggest specific next actions.
Systems such as OmniActions~\cite{li2024omniactions}, ContextAgent~\cite{yang2025contextagent}, Sensible Agent~\cite{lee2025sensible}, and Satori~\cite{li2024satori} predict or proactively suggest \textit{single, immediate actions}. 
For example, when a user is shopping for clothes and viewing a pair of jeans, such systems might suggest actions like \emph{``look up reviews,''} \emph{``search for similar items,''} or \emph{``check price comparisons''}.
These approaches solve a different problem.
Our setting asks not what the user should do next, but what \emph{set of functionalities} should remain available throughout the activity. 
In the same scenario, \rv{the relevant set might} include browsing related products, viewing reviews and product details, comparing prices, messaging a friend for a second opinion, and controlling background music, \rv{all} surfaced as persistent, spatially arranged entry points. 
This shifts the problem from immediate action prediction to holistic support, where suggestion quality depends not only on the relevance of individual suggestions but also on the \emph{coverage} the set provides across user needs in context.

This difference in problem formulation also changes evaluation. 
Prior systems \rv{were typically} assessed by next-action prediction accuracy or user experience metrics such as task load and usability.
These fit immediate assistive actions and interaction techniques, but are less suited to systematically comparing broader functionality suggestions.
Our work addresses this gap through a benchmarking framework for context-aware XR functionality suggestion, enabling systematic, reproducible comparison of algorithmic approaches.

\subsection{Computational Approaches for Structured UI Development}
Outside XR, a substantial body of HCI research treats interface development as a computational problem, supported by large-scale datasets, structured representations, \rv{and well-defined tasks for} retrieval, evaluation, and adaptation.
Datasets such as Webzeitgeist~\cite{kumar2013webzeitgeist}, ERICA~\cite{deka2016erica}, RICO~\cite{deka2017rico}, CLAY~\cite{li2022learning}, and UEyes~\cite{jiang2023UEyes} have enabled data-driven analysis of how interfaces are composed, perceived, and used at scale. 
Building on such resources, researchers have developed representations that make UI structure computationally accessible, including hierarchical screen parsing~\cite{wu2021screen}, graph-based formulations~\cite{jiang2024graph4gui}, semantic embeddings~\cite{li2021screen2vec}, and text-like encodings of view hierarchies for LLM reasoning~\cite{wang2023enabling}.
Collectively, these efforts establish interfaces as structured computational objects rather than monolithic visual artifacts.

This structured understanding has enabled downstream work on retrieval and automated evaluation. 
Embedding-based approaches support retrieval of semantically related screens, components, and commands from screenshots~\cite{li2021screen2vec,bai2021UIBerta}, sketches~\cite{huang2019swire}, layout structure~\cite{bunian2021vins}, or natural-language descriptions~\cite{adar2014commandspace,wang2021screen2words}, while other systems use learned structure to score design quality~\cite{wu2024uiclip,duan2024UICrit} and reconstruct interfaces across devices and form factors~\cite{wu2022reflow}.
More broadly, this line of work shows the value of explicit representations, reusable datasets, and well-defined computational tasks for making interface development more systematic and comparable.
\rv{We} extend this agenda to XR, where the challenge is not only to represent or adapt interfaces, but to determine \rv{which} task-relevant functionality to surface across a spatially distributed interface configuration.

\subsection{Adaptive XR Interfaces}
\rv{XR interfaces combine demands that prior work has largely studied separately.
Desktop and large-display systems support access to many functions across windows and displays~(\eg\cite{lischke2016screen,vogel2004interactive}), while mobile and ubiquitous computing adapt interfaces to users' changing situations~(\eg\cite{dey1999towards,baldauf2007survey}).
XR extends both challenges: interfaces span many applications and windows distributed throughout the surrounding environment, while their relevance shifts continuously across tasks, activities, and settings~\cite{Cheng25arproductivity,grubert2017towards,todi2023framework,davari2024contextawareadaptationextendedreality}.
Adaptation is therefore not merely an enhancement, but a necessary means of managing XR's expanded interface space, growing functional complexity, and continuously changing context~\cite{lindlbauer2022future}.
}

Prior work has 
\rv{explored a range of contextual signals for driving such adaptation, including users'} cognitive load~\cite{lindlbauer2019context}, individual preferences~\cite{cho2024minexr,johns2023pareto,johns2023towards}, and the semantics and geometry~\cite{cheng2021semanticadapt,cheng2023interactionadapt,han2023blendmr,fender2017heatspace} of the environment.
\rv{More recent systems have also prompted vision-language models to infer relevant situational signals from the user's surroundings~\cite{li2026Automating,li2024situationadapt}}.

\rv{Given these signals, XR interfaces may adapt their} \emph{contents} (\ie what is available and how much), \emph{presentation} (\ie how and when it is shown), and \emph{placement} (\ie where it appears)~\cite{todi2023framework}.
Much existing work has concentrated on placement and visual configuration: how windows and interface elements should be arranged in space or adjusted to environmental constraints.
This focus is well motivated, since XR removes many of the spatial restrictions of conventional UIs and makes layout itself a central design problem.
However, \rv{this work typically assumes} the functionality to present is already given, \rv{or selects it} only at \rv{the granularity of whole applications.}
Our work complements this line of research by \rv{addressing \emph{content}: which functional facets} should be surfaced in the first place\rv{, before deciding} how or where to present them.

\section{\system{}: Benchmarking Framework}
\system{} is a benchmarking framework for evaluating context-aware XR systems across users, tasks, and environments. 
It enables researchers to answer questions such as which contextual signals are most informative, which modeling strategies perform best, and how much \cost{} is reduced through assistance. 
The framework consists of (1) a functionality-centric representation based on \textbf{functional facets}; (2) \textbf{\textsc{MineXR++}}, an augmented dataset capturing facet usage across contexts; (3) \textbf{problem formulations} for context-aware functional suggestion; and (4) an evaluation protocol centered on \textbf{\cost{}}.

% graph-based UI representations~\cite{wu2021screen,jiang2024graph4gui}
% semantic UI representations~\cite{li2021screen2vec,wang2023Enabling}
% action suggestion systems~\cite{li2024omniactions,yang2025contextagent,lee2025sensible,li2024satori}
% findings from MineXR~\cite{cho2024minexr}
% tool-based agents~\cite{nakano2021webgpt,yao2023react,schick2023toolformer}

\subsection{Representation: Functional Facets}
Our framework uses the functional facet as its central unit of computation.

\paragraph{Why a new representation?}
Context-aware XR interfaces must decide \emph{what level of functionality} to surface in a limited spatial interface.
If the unit is too coarse, such as an entire application, the interface exposes more functionality than the current task and environment require, increasing visual clutter and search effort.
If the unit is too fine, such as individual atomic actions or API calls, the interface becomes fragmented and forces users to repeatedly assemble low-level controls to accomplish a single goal.

Applications are naturally composed of many atomic actions.
For example, \rv{a music app} exposes actions like \texttt{play}, \texttt{pause}, and \texttt{shuffle}, and a recipe app exposes actions like \texttt{view\_ingredients}, \texttt{view\_preparation\_steps}, and \texttt{view\_cooking\_time}.
While atomic actions are suitable for direct command-style interaction (\eg \textit{``Stop the music.''} or \textit{``How long does cooking this recipe take?''}), they are often too fine-grained for persistent, visually surfaced XR interfaces.
\rv{Surfacing capabilities in isolation increases interaction effort, forcing users to reassemble the related capabilities they need.}
% Showing isolated controls without their related functionality can increase interaction effort by requiring additional browsing, speech commands, or manual reconfiguration to reach the desired functionality.

This \rv{motivates} an intermediate representation that is more specific than a monolithic application but more usable than individual actions.
Our framework addresses this need through \emph{functional facets}.
\rv{In designing functional facets, we drew on several lines of prior work. 
Graph-based UI representations, which model interfaces as nodes and transitions, inform the facet graph~\cite{wu2021screen,jiang2024graph4gui}. Semantic UI representations, which encode the meaning of interface content rather than its structural UI form, inform the intent-level naming of facets~\cite{li2021screen2vec,wang2023enabling}.
In addition, action-suggestion systems~\cite{li2024omniactions,yang2025contextagent,lee2025sensible,li2024satori} and tool abstractions used by LLM-based agents~\cite{nakano2021webgpt,yao2023react,schick2023toolformer} inform the grounding of capabilities in discrete, invocable units.
The specific facet inventory is derived from MineXR~\cite{cho2024minexr}, whose envisioned XR
layouts ground the representation in functionality that users actually wanted
surfaced in context.
}

\begin{figure*}
    \centering
    \includegraphics[width=0.9\linewidth]{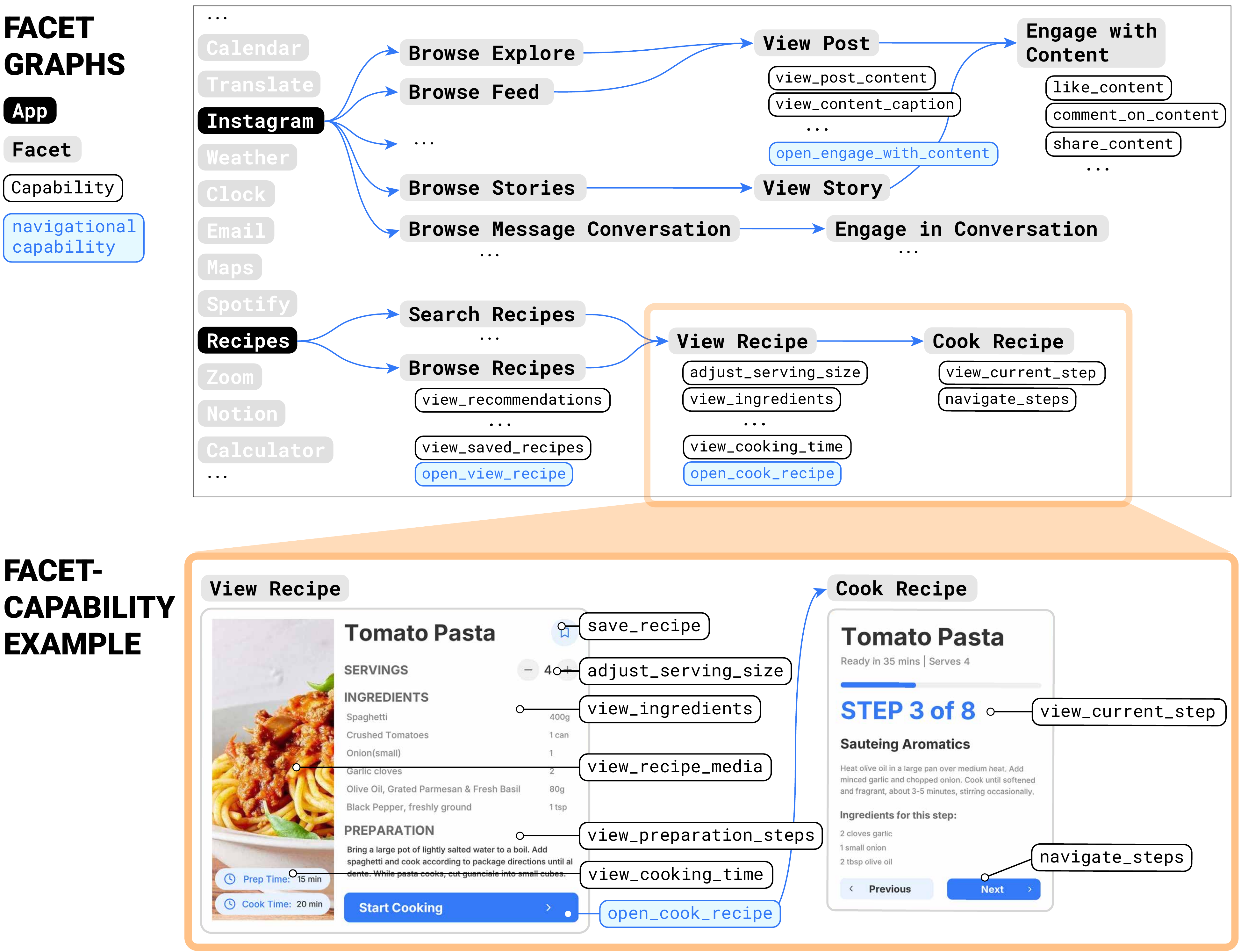}
    \caption{Top: Snippet of app--facet--capability structure. Bottom: Visualized example of facets and their component capabilities. A navigational capability (\eg \texttt{open\_cook\_recipe}) leads to another facet (\eg \texttt{Cook Recipe}). }
    \label{fig:facet-example}
\end{figure*}

\subsubsection{Functional facets}
A \emph{functional facet} is a semantically coherent grouping of one or more related atomic actions, which we call \emph{capabilities}, that together support a shared \emph{user intent}.
Each app $a \in \mathcal{A}$ \rv{has} a set of facets $\mathcal{F}_a$.
We represent a facet $f \in \mathcal{F}_a$ as
\[
f = (i_f, \mathcal{K}_f),
\] 
where $i_f$ \rv{is the facet's} user intent and $\mathcal{K}_f$ \rv{is its} set of capabilities.\
% each app $a \in \mathcal{A}$ is associated with a set of facets $\mathcal{F}_a$, and each facet $f \in \mathcal{F}_a$ has a user intent $i_f$ and a corresponding set of capabilities $\mathcal{K}_f$.
Facet names \rv{express} this intent in a \texttt{verb+noun} format.
For example, the facet \texttt{View Recipe}~\rv{(\autoref{fig:facet-example})} serves the intent of viewing a recipe, supported by capabilities such as \texttt{view\_ingredients}, \texttt{view\_preparation\_steps}, 
 and \texttt{view\_cooking\_time}.
Similarly, in a music app, the facet \texttt{Control Playback} is supported by capabilities such as \texttt{play}, \texttt{pause}, \texttt{skip\_next}, and \texttt{shuffle}.

\subsubsection{Facet graphs}
Facets of the same application are inherently connected through the underlying capabilities.
We model these connections using a facet graph 
\[
G_a = (\mathcal{F}_a, \mathcal{E}_a),
\]
where \rv{a directed edge} $(f_u, f_v) \in \mathcal{E}_a$ \rv{exists} if facet $f_u$ contains a navigational capability that opens facet $f_v$.
% capabilities \rv{can be connected through} directed edges between facets:
% \[
% G_a = (\mathcal{F}_a, \mathcal{E}_a),
% \]
% where $(f_u, f_v) \in \mathcal{E}_a$ if facet $f_u$ contains a navigational capability that opens facet $f_v$.
\emph{Navigational capabilities} enable transitions between facets, while other capabilities \rv{in $\mathcal{K}_f$} provide information or actions within a facet.
Navigational capabilities follow an \texttt{open\_<facet\_name>} convention, reflecting the intent to open the target facet.

\autoref{fig:facet-example} illustrates an example.
The black nodes \rv{labeled} with app names represent a special \texttt{Open Section} facet, corresponding to each app's entry point (\eg the app icon) or central navigation hub (\eg a navigation bar) \rv{that links} to \rv{its} main facets.
For instance, Instagram connects to \texttt{Browse Explore}, \texttt{Browse Feed}, \texttt{Browse Stories}, and \texttt{Browse Message Conversation}, while Recipes connects to \texttt{Search Recipes} and \texttt{Browse Recipes}.
\rv{Blue nodes denote navigational capabilities: the edge} \texttt{open\_view\_recipe}, for example, connects \texttt{Browse Recipes} to \texttt{View Recipe}.

The lower portion of \autoref{fig:facet-example} grounds this representation in a concrete interface. 
The \texttt{View Recipe} facet is instantiated as a recipe page, where capabilities such as \texttt{view\_ingredients} and \texttt{view\_cooking\_time} correspond to visible UI elements.
The ``Start Cooking'' button implements \texttt{open\_cook\_recipe}, triggering a transition to the \texttt{Cook Recipe} facet, which \rv{guides the user through} step-by-step instructions \rv{via} \texttt{view\_current\_step} and \texttt{navigate\_steps}.

\subsection{Dataset: MineXR++}
We instantiate this computational framework with \textsc{MineXR++}, an augmented version of the open-source MineXR dataset~\cite{cho2024minexr} annotated with the functional facet structure.

MineXR contains 109 personalized XR layouts and 695 widgets from 31 participants across everyday activities such as focus work, chatting with friends, cooking, and relaxing in four environments: office, living room, kitchen, and coffee shop.
\rv{Participants created these layouts by taking} screenshots from apps \rv{and websites, or drawing sketches,} and cropping the regions they wanted to place in the environment as custom widgets.

While MineXR offers a valuable view into envisioned everyday XR use, its original form is not directly suitable for computational evaluation.
Widgets are free-form image crops, and the accompanying annotations do not provide a consistent structure for representing functionality across apps and scenarios.

To \rv{operationalize} the dataset within our framework, we augmented MineXR with a functional-facet annotation layer.
We first constructed a functional facet database.
Facet definitions were derived from MineXR screenshots and additional screenshots from recent app versions, and organized around recurring user intents and their associated capabilities.
To reduce sparsity, applications with highly similar functionality and limited usage were consolidated into shared categories.
\rv{The resulting database covers} 42 applications, comprising 273 facets and 1,007 capabilities\rv{, of which} 101 facets were placed by participants during the MineXR data collection.

Using this database, we annotated each widget with its corresponding functional facet(s) and the specific capabilities it exposed in context.
Annotation criteria were iteratively discussed among three authors; primary annotation was conducted by one author and reviewed by a second author.
We refer to this facet-annotated version as \textsc{MineXR++}.

\begin{figure}[t]
    \centering
    \includegraphics[width=1\linewidth]{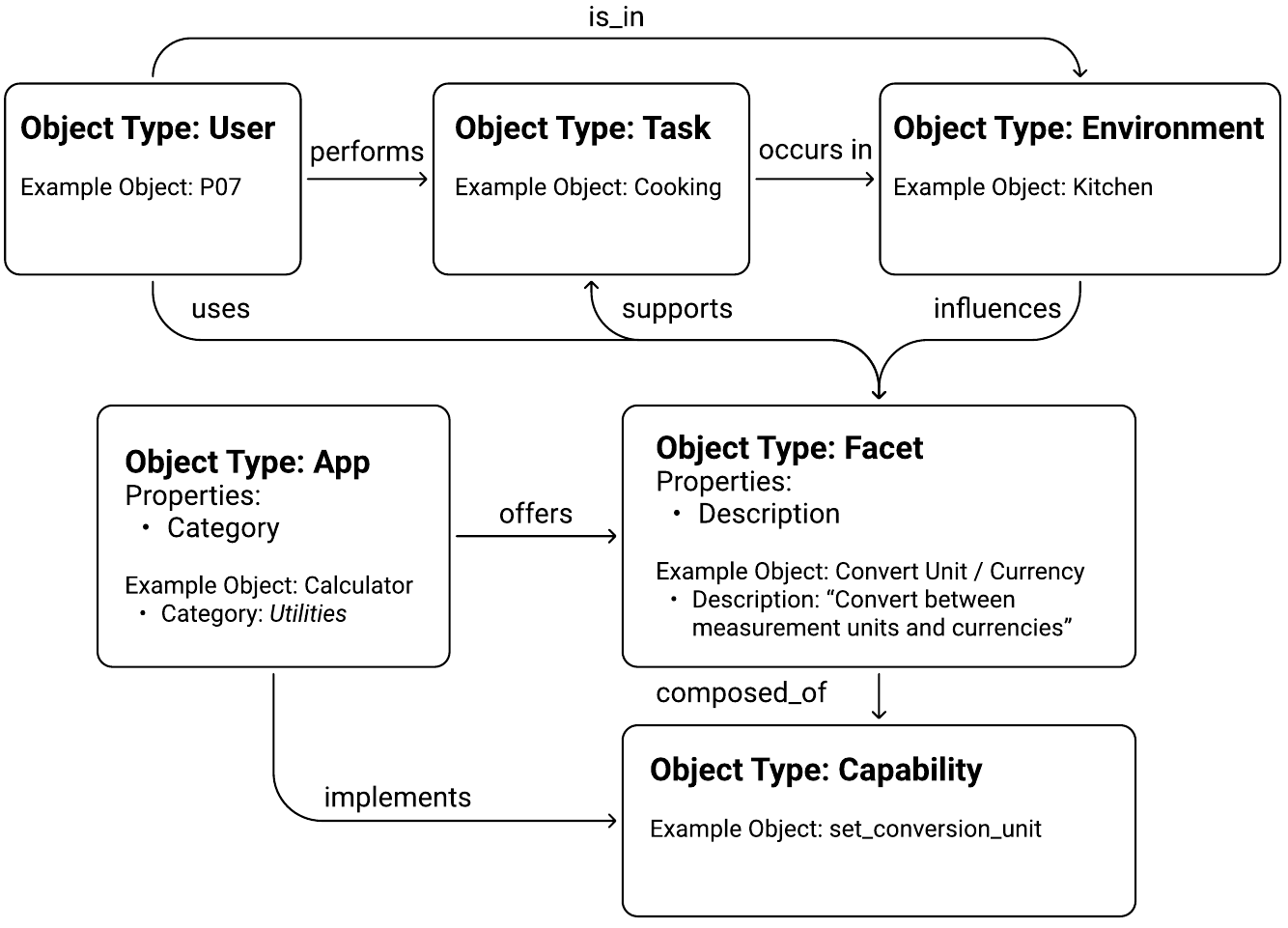}
    \caption{\rv{Object} schema \rv{relating functional facets to their app, capabilities, and context of use (user, task, environment). This schema defines the structure of \textsc{MineXR++} and is traversed by the relational retrieval method in Section~\ref{sec:experiments}.}}
    \label{fig:relational-schema}
\end{figure}

In summary, each facet contains information about what app it originates in and which capabilities it is composed of.
\rv{Each placed widget} in \textsc{MineXR++} \rv{links a facet to its context of use:} user ID (\eg ``P07''), task (\eg ``cooking''), and location (\eg ``kitchen''), as shown in the schema in \autoref{fig:relational-schema}\rv{, which also serves as the structure our relational retrieval method traverses~(Section~\ref{sec:experiments})}.
\rv{These fields} are the minimum requirements for a dataset to be compatible with our proposed benchmarking framework, which in the future could be augmented with information about placement, grouping, user state (\eg cognitive load) or user capabilities. 

\subsection{Problem Formulation}
We formulate \problem{} as the problem of deciding \emph{which functional facets to make readily accessible} for a given usage context.
The goal is not to predict a single next action, but to identify the subset of functionality that should be surfaced so that users can reach what they need with low friction.

\rv{Our benchmark operationalizes context as a triple of user identity $u$, task $t$, and environment $e$, following the structure of MineXR~\cite{cho2024minexr}.
This lets us study how these factors relate to the functionality a user needs.}
For a given context $q=(u,t,e)$, the desired outcome is a target set of functional facets, $Y_q \subseteq \mathcal{F}$, where $\mathcal{F}$ is the set of candidate facets across applications.
Intuitively, $Y_q$ \rv{is} the functionality that should be easy to reach in that situation.
In \textsc{MineXR++}, $Y_q$ is the set of facets annotated on the XR elements that participants placed in their envisioned XR layouts.
A prediction method produces a subset $\hat{Y}_q \subseteq \mathcal{F}$, the facets it suggests for that context\rv{, which we evaluate against the ground-truth target $Y_q$}.
% In other words, $Y_q $ represents the ground truth and $\hat{Y}_q$ is a predicted set of facets.

This formulation is agnostic to how $\hat{Y}_q$ is produced: a system may retrieve it from similar examples, infer it from language or structured context, generate it with an LLM, or combine multiple strategies.
What matters for the framework is that different methods can be evaluated on the same underlying task, surfacing the right functional facets for the right context.

\subsection{Evaluation Protocol: Simulated Interaction}
To evaluate suggestions, \system{} models the \textbf{\cost{}} required to access the desired facet set $Y_q$ given a suggested set $\hat{Y}_q$.

\paragraph{Why \cost{}?}
To evaluate \problem{}, we aim to measure how much effort suggestions save a user in reaching the desired set of functional facets, rather than relying solely on set overlap with ground-truth facets.
In XR, usefulness depends not only on exact correctness but on how suggestions reduce navigation and search effort.
For example, a closely related facet from the correct application may still be substantially more useful than an unrelated one, even if both would be counted as incorrect under exact-match metrics.
Crucially, this requires evaluating \emph{sets} of suggested functionality: systems surface multiple facets at once, and users can choose among them, so utility depends on how the \emph{entire set} supports access.
\rv{Whereas} top-k metrics such as Recall@K or Hit@K reduce predictions to binary correctness at a cutoff, \cost{} captures partial relevance and how a set of suggestions jointly reduces user effort, and enables direct comparison against an unassisted manual baseline.
\rv{In taking access effort rather than prediction accuracy as the unit of
evaluation, we follow menu-optimization work that scores interface designs by the
time or steps a user spends reaching a target~\cite{todi2021adapting,bailly2013menuoptimizer}.
We adapt this idea to functional suggestion, using \cost{} as a complement to the
retrieval and prediction metrics above rather than a replacement for them.}

\subsubsection{Navigation \& search cost} \label{sec:interaction-cost}
The \cost{} model measures the effort a user \rv{spends reaching} the desired set of functional facets, following a simple simulated user strategy.
\rv{It is a deliberately simplified model of access: a user opens a menu, scans for the relevant application or suggestion, and navigates to the target facet, with each step assigned a fixed cost. 
This abstracts away much of the perceptual, motor, and cognitive variation of real XR interaction, and is intended for comparing methods under identical assumptions rather than estimating absolute human performance. 
The framework is designed so that this layer can be replaced with richer cost models, discussed in Section~\ref{sec:discussion-extending}.}

\paragraph{Manual baseline.}
\rv{Without system support}, the user opens the application menu and navigates to each desired facet without any system support.
The total manual cost is
\[
C_{\text{manual}}(Y_q)
=
c_{\text{open\_menu}}
+
\sum_{f_i \in Y_q}
\left(
c_{\text{app\_scan}}(f_i)
+
c_{\text{facet\_nav}}(f_i)
\right),
\]
\sloppy{\rv{where} $c_{\text{open\_menu}}$ is the one-time cost of opening the app list, $c_{\text{app\_scan}}(f_i)$ is the cost of locating the relevant application for facet $f_i$, and $c_{\text{facet\_nav}}(f_i)$ is the cost of navigating within that application $G_a$ to reach $f_i$.}

\paragraph{Suggestion-assisted access.}
\rv{When a system suggests $\hat{Y}_q$,}
% Considering scenarios that include system support for providing access to facets, 
each target facet in $Y_q$ is resolved into one of three disjoint subsets:
\begin{itemize}
    \item $H$: exact hits, where the desired facet itself is directly surfaced
    \item $A$: same-app adjustments, where a surfaced facet from the same application provides a useful entry point but is not itself the desired facet
    \item $M$: misses, where no helpful suggestion is available and the user must fall back to manual access.
\end{itemize}
These subsets \rv{partition the target set:} $Y_q = H \cup A \cup M$.
% form the complete list of suggestions presented by a system, denoted as 

Exact hits are directly accessible and incur no additional navigation cost.
For same-app adjustments, we assume that the surfaced facet places the user within the correct application, so the remaining effort is smaller than full manual access but still non-zero.
This residual effort is captured by an adjustment cost.
\rv{Misses require falling back to manual navigation, and the menu is opened only when at least one miss occurs.}

The suggestion-assisted cost is
\[
\begin{aligned}
&C_{\text{suggest}}(Y_q,\hat{Y}_q)
=
c_{\text{suggest\_scan}}(\hat{Y}_q)
+
\sum_{f_i \in A} c_{\text{adjust}}(f_i) \\
&\quad +
\sum_{f_i \in M}
\left(
c_{\text{app\_scan}}(f_i)
+
c_{\text{facet\_nav}}(f_i)
\right)
+
\mathbf{1}_{|M|>0}\, c_{\text{open\_menu}},
\end{aligned}
\]
where $c_{\text{suggest\_scan}}(\hat{Y}_q)$ is the cost of scanning the surfaced suggestions, $c_{\text{adjust}}(f_i)$ is the residual cost of reaching $f_i$ from a same-app suggestion, and the final terms capture fallback to manual access for misses.
A lower \cost{} indicates that the surfaced subset provides more efficient access to the desired functionality.

\subsubsection{Supplementary retrieval metrics.}
To support comparison with standard retrieval and recommendation settings, the framework can also report traditional metrics such as Recall@K, F1@K, and Hit@K.
We treat these as complementary measures of retrieval-style correctness, while \cost{} captures user-facing access efficiency.

\section{Experiments} \label{sec:experiments}
We \rv{apply} \system{} in three experiments\rv{, each targeting a canonical question that arises when developing a context-aware XR system}: (1) a \textbf{context factor analysis} \rv{asking \emph{which contextual signals matter}---how much a user's identity, task, and environment each explains the functionality they need;} 
(2) an \textbf{initial facet suggestion} task \rv{asking \emph{what to surface when a user first enters a context}, before any interaction has unfolded};
and (3) a \textbf{next facet suggestion} task \rv{asking \emph{what to surface next as the interaction progresses}, using the functionality already in use as additional evidence.}
\rv{These illustrate how \system{} can support systematic, reproducible evaluation, spanning the understanding of context, cold entry into a context, and ongoing interaction within it.}

\subsection{Experiment 1: Context Factor Analysis} \label{sec:lift-analysis}
\subsubsection{\rv{Task}}
\rv{Before deciding what to surface, a developer needs to know which context factors actually shape a user's needs: does desired functionality track \emph{who} the user is, \emph{what} they are doing, \emph{where} they are, or some combination?
This experiment quantifies, for each facet, which of the three context factors (user, task, environment) most strongly predicts its use, using the human-authored target sets in \textsc{MineXR++} as ground truth.}

\subsubsection{\rv{Methods and procedure}}
To examine how desired functionality depends on context, we performed a \emph{lift}-based analysis~\cite{tuffry2011datamining} at the facet level.
For a given context factor, lift compares how often a facet appears under a matched condition (for example, the same user, task, or environment) relative to how often it appears overall in the dataset.
A lift greater than 1 indicates that the facet is more likely than expected under that condition.
For each facet, we identified which context factor produced the strongest lift, \ie \rv{the factor most associated with its use}. 
% factors are dominant in determining how often a facet will be used), yielding a mixed overall pattern illustrated on the left side of \autoref{fig:lift-analysis}.

\subsubsection{\rv{Results}}
\begin{figure}[t]
    \centering
    \includegraphics[width=\linewidth]{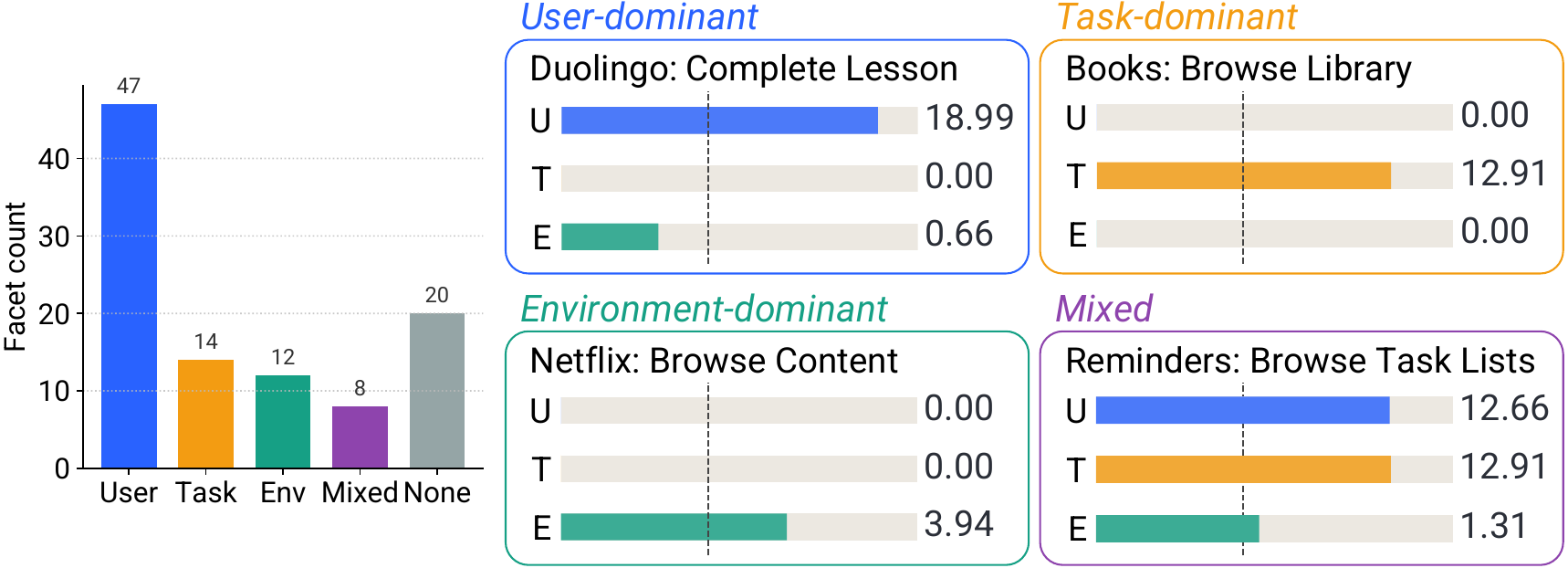}
    \caption{\rv{Left: distribution of the dominant context factor across facets (user-, task-, environment-dominant, mixed, or no lift). Right: one representative facet per pattern, showing lift by context factor.}}
    % Dominant context factor per facet based on lift over baseline prevalence, with one example each for user-, task-, environment-, and mixed-dominant patterns.}
    \label{fig:lift-analysis}
\end{figure}

\rv{The left of \autoref{fig:lift-analysis} shows the distribution of dominant factors across facets.}
Among facets used in \textsc{MineXR++}, 47 were user-dominant, 14 task-dominant, and 12 environment-dominant, while 8 showed mixed dependence and 20 showed no lift above baseline.
This variability suggests that desired functionality in XR \rv{does not reduce} to a single factor alone, but reflects a combination of personal preference, task relevance, and environment setting.

The example lift profiles on the right side of \autoref{fig:lift-analysis} show one representative facet for each pattern. 
\texttt{Duolingo: Complete Lesson} is user-dominant, \texttt{Books: Browse Library} is task-dominant, and \texttt{Netflix: Browse Content} is environment-dominant. \texttt{Reminders: Browse Task Lists} shows a mixed profile, with similarly strong user and task lift.
These examples \rv{indicate} that different facets align with different contextual axes, motivating \rv{suggestion} methods that can incorporate different kinds of prior evidence.
\rv{The facet} representation \rv{makes this} analysis \rv{possible, offering a way to examine which factors drive functional needs in} context-aware XR.

\subsection{Experiment 2: Initial Facet Suggestion}
\subsubsection{\rv{Task}}
When a user first enters a context $q$, \rv{an} XR system \rv{must} decide what functionality to surface before any interaction unfolds.
We frame this as \emph{initial facet suggestion}: given the current user, task, and environment, predict a set of facets $\hat{Y}_q$ \rv{to make} readily accessible at the start of the episode.
\rv{The target $Y_q$ is the set of facets the participant placed for that context
in \textsc{MineXR++}, so a method is scored on how well it recovers a real user's
situated selection.}
This setting captures cold-start and example-augmented support, and evaluates how far different methods reduce the effort required to reach desired functionality relative to manual access.

\begin{table*}[t]
\centering
\begin{minipage}[t]{0.48\textwidth}
\centering
\small
\setlength{\tabcolsep}{4pt}
\renewcommand{\arraystretch}{1.12}
% \begin{threeparttable}
\caption{Average \cost{} and retrieval quality for \emph{initial facet suggestion}. Parentheses indicate the number of evaluated scenarios. Cost reductions are relative to manual access computed on the same scenarios.}
\label{tab:init-suggestion-results}
\begin{tabular}{@{}lcccc@{}}
\toprule
\textbf{Condition} & \textbf{Cost} & \textbf{Recall@10} & \textbf{F1@10} & \textbf{Hit@10} \\
\midrule
\textbf{Manual Access} & 21.20 & -- & -- & -- \\
\midrule
\rowcolor[gray]{0.95}
\textbf{Global Popularity} & 19.85 \costdrop{(-6.4\%)} & 0.167 & 0.132 & 0.688 \\
\textbf{Relational Retrieval} & 18.39 \costdrop{(-13.3\%)} & 0.259 & 0.193 & 0.798 \\
\rowcolor[gray]{0.95}
\textbf{Zero-shot LLM} & 17.66 \costdrop{(-16.7\%)} & 0.228 & 0.168 & 0.761 \\
\multicolumn{5}{@{}l}{\textbf{Few-shot LLM}} \\
\quad User \textit{(106)} & 16.33 \costdrop{(-22.6\%)} & \textbf{0.338} & \textbf{0.255} & 0.830 \\
\quad Task \textit{(40)} & 18.41 \costdrop{(-15.9\%)} & 0.274 & 0.217 & 0.825 \\
\quad Environment \textit{(108)} & 17.07 \costdrop{(-18.8\%)} & 0.287 & 0.206 & \textbf{0.870} \\
\addlinespace[2pt]
\quad User+Task \textit{(\textcolor{red}{2})} & \textbf{15.25} \costdrop{\textbf{(-30.7\%)}} & \textbf{0.464} & \textbf{0.399} & \textbf{1.000} \\
\quad User+Env \textit{(94)} & \textbf{15.53} \costdrop{\textbf{(-22.7\%)}} & 0.319 & 0.229 & 0.862 \\
\quad Task+Env \textit{(47)} & 19.54 \costdrop{(-12.6\%)} & 0.222 & 0.178 & 0.702 \\
\bottomrule
\end{tabular}
% \end{threeparttable}
\end{minipage}
\hfill
\begin{minipage}[t]{0.48\textwidth}
\centering
\small
\setlength{\tabcolsep}{4pt}
\renewcommand{\arraystretch}{1.12}
% \begin{threeparttable}
\caption{Average \cost{} and retrieval quality for \emph{next facet suggestion}. Scenarios are weighted equally after averaging over their timelines. Cost reductions are relative to manual access computed on the same scenarios.}
\label{tab:next-facet-summary}
\begin{tabular}{@{}lcccc@{}}
\toprule
\textbf{Condition} & \textbf{Cost} & \textbf{Recall@10} & \textbf{F1@10} & \textbf{Hit@10} \\
\midrule
\textbf{Manual Access} & 21.19 & -- & -- & -- \\
\midrule
\rowcolor[gray]{0.95}
\textbf{Global Popularity} & 18.70 \costdrop{(-11.7\%)} & 0.168 & 0.089 & 0.466 \\
\textbf{Relational Retrieval} & 15.70 \costdrop{(-25.9\%)} & \textbf{0.355} & \textbf{0.162} & \textbf{0.687} \\
\rowcolor[gray]{0.95}
\textbf{Zero-shot LLM} & 16.00 \costdrop{(-24.5\%)} & 0.205 & 0.102 & 0.488 \\
\multicolumn{5}{@{}l}{\textbf{Few-shot LLM}} \\
\quad User & \textbf{14.61} \costdrop{\textbf{(-31.1\%)}} & 0.285 & 0.147 & 0.574 \\
\quad Task & 19.22 \costdrop{(-9.3\%)} & 0.079 & 0.044 & 0.188 \\
\quad Environment & 15.82 \costdrop{(-25.3\%)} & 0.209 & 0.104 & 0.492 \\
\addlinespace[2pt]
\quad User+Task & 20.98 \costdrop{(-1.0\%)} & 0.007 & 0.004 & 0.014 \\
\quad User+Env & 15.83 \costdrop{(-25.3\%)} & 0.257 & 0.122 & 0.494 \\
\quad Task+Env & 19.30 \costdrop{(-8.9\%)} & 0.090 & 0.046 & 0.205 \\
\bottomrule
\end{tabular}
% \end{threeparttable}
\end{minipage}

\end{table*}

\subsubsection{Methods}
We use \system{} to compare five strategies: manual access, global popularity, relational retrieval, zero-shot LLM, and few-shot LLM.
\rv{These span context-agnostic baselines, structured retrieval, and generative reasoning, chosen to represent the method families used in adaptive XR rather than to reimplement any single prior system.
We include the LLM conditions to represent the recent class of context-aware assistance systems that use large language models to reason over sensed context~\cite{li2024omniactions,lee2025sensible,yang2025contextagent}. 
Because those systems predict single next actions rather than sets of persistent functionality (Section~\ref{sec:related}), we do not compare against them directly; instead, our LLM baselines apply the same underlying contextual-reasoning capability to our set-based task, so that generative and non-generative approaches can be compared under one benchmark.
}
For all assisted methods, the system returns a top-$10$ set of suggested facets, evaluated with both \cost{} and standard retrieval metrics.

\paragraph{Manual access.}
Manual access is the unassisted reference condition: no facets are proactively suggested, and the user reaches desired functionality through manual navigation.
This is computed using the simulated interaction procedure of Section~\ref{sec:interaction-cost}.

\paragraph{Global popularity.}
As a simple low-cost baseline, global popularity ignores the query context and ranks facets only by how often they appear in the training data.
This provides a context-agnostic prior over commonly needed functionality.

\paragraph{Relational retrieval.}
Relational retrieval uses a structured context--facet representation~(\autoref{fig:relational-schema}).
Each dataset entry is a context--facet event, from which we construct a typed entity--relation schema with nodes for users, tasks, environments, apps, facets, and capabilities, linked by fixed relations such as user-\textit{performs}-task, user-\textit{uses}-facet, and facet-\textit{composed-of}-capability.
The schema provides structured retrieval by defining the entities and relations over which counts, adjacency maps, and capability links are computed.
Given a query \rv{context} $q=(u,t,e)$, the model ranks candidate facets using context-conditioned statistics from the training data, combining signals such as $P(f|u)$, $P(f|t)$, $P(f|e)$, $P(f|u,t)$, $P(f |u,e)$, and $P(f|t,e)$.
To remain robust when exact matches are sparse, it also \rv{applies} soft transfer over task and environment labels using cosine similarity between Sentence-BERT embeddings from \texttt{all-MiniLM-L6-v2}~\cite{reimers2019sentence}.

This method is motivated in part by XR deployment constraints. 
Compared with LLM-based inference, retrieval over precomputed relational statistics is substantially cheaper and lower-latency, \rv{offering a favorable accuracy--cost tradeoff} for resource-constrained headsets, where compute and battery use directly affect usability.

\paragraph{LLM-based suggestion.}
\rv{The two LLM conditions} operate over a structured prompt~\rv{(\autoref{sec:prompts})} containing the current scenario (task and environment) and a catalog of candidate facets with descriptions, capabilities, and facet-graph links.

The \textbf{zero-shot LLM} predicts relevant facets from the current context and catalog \rv{alone}, without retrieved examples.
The \textbf{few-shot LLM} extends this through in-context learning~\cite{brown2020language}, adding historical scenarios and their facet selections under controlled transfer conditions (\eg matching user, task, or environment).

In both cases, the model selects a fixed-size set of facets (top-10), balancing direct task support, complementary functionality, and ambient utilities, and using the facet graph to reason about granularity and avoid redundant navigation chains.
% The few-shot variant uses examples as weak priors to capture recurring patterns in facet usage and composition.
All LLM results use Gemini~3~Flash at temperature 0.2.

\subsubsection{\rv{Procedure}}
\rv{Each method is evaluated on every eligible context $q$ in \textsc{MineXR++}, producing a top-10 suggested set $\hat{Y}_q$ that we compare against the human-authored target $Y_q$.
Global popularity and relational retrieval use seeded five-fold cross-validation at the scenario level, so that all observations from a held-out scenario are excluded from its training fold. 
For the few-shot conditions, we report results separately for each transfer condition (\eg \emph{User}, \emph{Task}, \emph{User+Env}), drawing historical examples only from scenarios matching the specified factor(s); parentheses in \autoref{tab:init-suggestion-results} give the number of evaluated scenarios for which such matched examples exist.
We report \cost{} as the primary measure, expressed as the percentage reduction relative to manual access on the same scenarios, alongside standard top-10 retrieval metrics (Recall@10, F1@10, Hit@10) as complementary measures of
retrieval-style correctness.}

\subsubsection{Results}
\autoref{tab:init-suggestion-results} summarizes the results.
All assisted methods reduce \cost{} relative to manual access, showing that even simple proactive support can make desired functionality easier to reach, though the benefit varies by suggestion strategy and by what prior information is available.

Global popularity yields only a modest improvement (\(-6.4\%\)), suggesting that frequently used facets alone \rv{do not} characterize what should be surfaced in a specific context.
Relational retrieval improves more substantially (\(-13.3\%\)), indicating that structured, context-conditioned historical signals \rv{capture more of the context-to-functionality mapping than frequency alone.}
Zero-shot LLM performs better still (\(-16.7\%\)), implying that general contextual reasoning already captures part of the mapping \rv{without task-specific statistics}.

The strongest results come from few-shot LLM conditions, though their effectiveness depends strongly on which context factors are matched.
\textit{User+Task} is best numerically, but with only two matched scenarios we interpret it cautiously. 
Among more broadly supported settings, \textit{User} and \textit{User+Env} perform best, reducing \cost{} by \(-22.6\%\) and \(-22.7\%\) \rv{respectively}.
This suggests that user-aligned examples are especially valuable for initial facet suggestion,
a pattern partially consistent with the lift analysis (Section~\ref{sec:lift-analysis}), where many facets showed strong user-dependent lift.
By contrast, \textit{Task+Env} underperforms the other pairwise settings, further suggesting that user-specific regularities \rv{matter most} for initial facet suggestion in our data.

\begin{figure}[t]
    \centering
    \includegraphics[width=\linewidth]{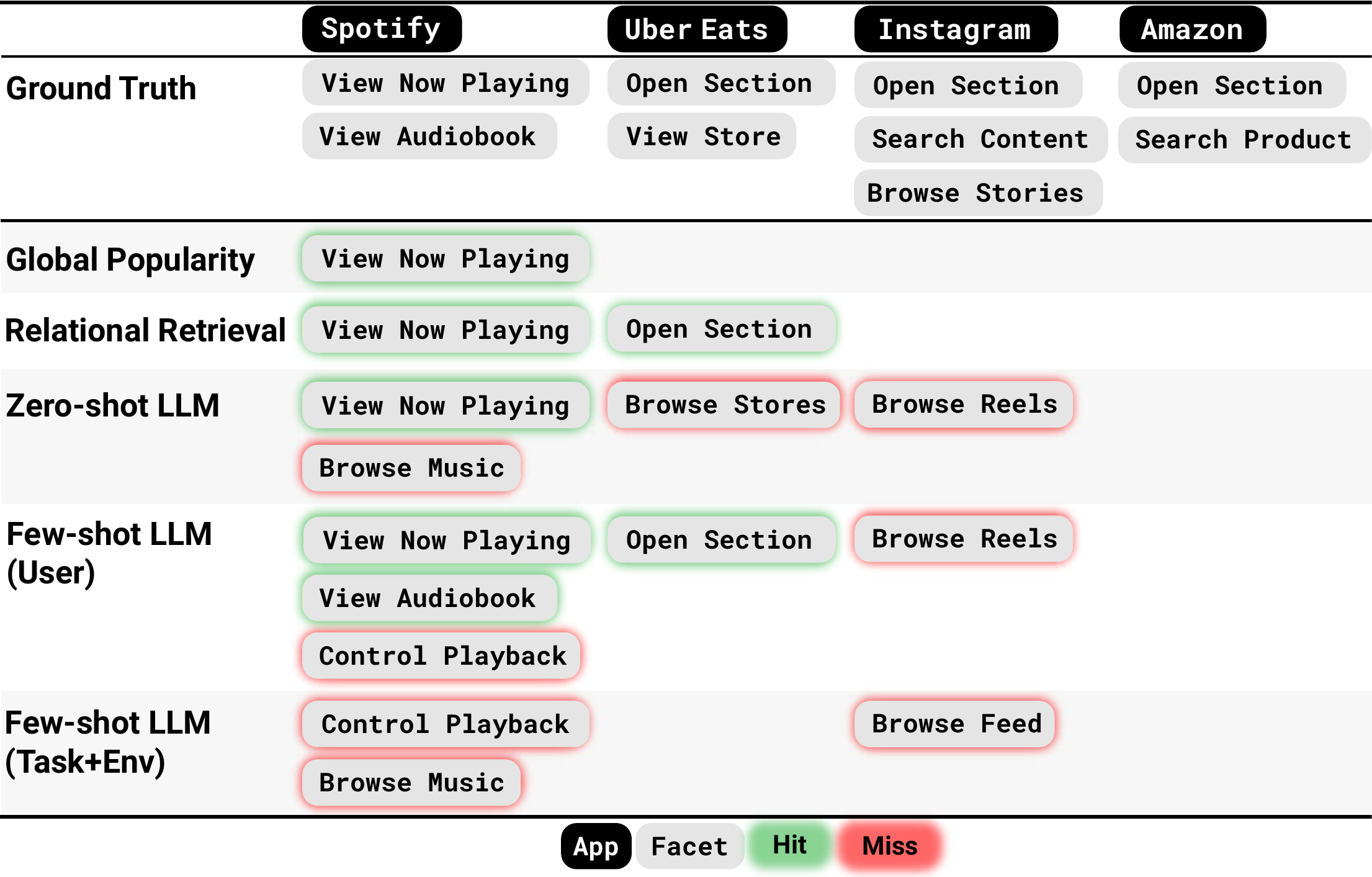}
    \caption{Example of initial facet suggestion for a specific participant and context (``P04'', ``relaxing'', ``living room''). By comparing ground-truth facets in \textsc{MineXR++} with suggestions from multiple methods, \system{} makes differences in functional coverage explicit for analysis.}
    \label{fig:qual_reslts}
    \vspace{-1em}
\end{figure}

\paragraph{Qualitative observations}
\autoref{fig:qual_reslts} illustrates a concrete example. For the context (``P04'', ``relaxing'', ``living room''), the ground truth includes nine facets from Spotify, Uber Eats, Instagram, and Amazon. 
Suggestions for other apps are omitted from this figure, and \texttt{Open Section} denotes an app's entry point or central navigation hub.

Global popularity surfaces only \texttt{Spotify: View Now Playing}, missing the relevant \texttt{Uber Eats}, \texttt{Instagram}, and \texttt{Amazon} facets. 
Relational retrieval adds \texttt{Uber Eats: Open Section} but still misses \texttt{Instagram} and \texttt{Amazon}.
Zero-shot LLM improves app-level grounding by suggesting the right apps, but often at the wrong facet granularity, \eg \texttt{Spotify: Browse Music} instead of \texttt{View Audiobook} and \texttt{Uber Eats: Browse Stores} instead of \texttt{Open Section}. 
\textit{User} few-shot prompting better captures this user's habitual choices, recovering both \texttt{Spotify: View Audiobook} and \texttt{View Now Playing} as well as \texttt{Uber Eats: Open Section}. 
\textit{Task+Env} examples are less helpful here, reflecting other users' preferences, such as \texttt{Netflix} and \texttt{Webtoon}, rather than this person's typical selections.

These \rv{observations} reinforce the lesson of Experiment~1: desired functionality in XR is shaped by multiple context factors, and methods that selectively incorporate the right prior evidence perform best.
They also \rv{show how} \system{} makes it possible to compare not only \emph{which model} performs better, but also \emph{which kinds of context information} are most useful for reducing \cost{} at the start of an interaction episode.
\rv{Using \system{},} new methods can be compared with existing \rv{ones} in a systematic, reproducible way.

\subsection{Experiment 3: Next Facet Suggestion}

\subsubsection{\rv{Task}}
As interaction unfolds, systems can leverage the facets already used in the current episode to infer what functionality is likely to be needed next.
In the \emph{next facet suggestion} task, the model predicts the next desired facet at each step of a scenario timeline, capturing step-by-step adaptive support rather than one-shot initialization.
\rv{Here the facets a user has already placed serve as additional contextual evidence: the task tests whether within-episode interaction history improves prediction beyond the static context $q=(u,t,e)$ alone.}

\subsubsection{\rv{Methods}}
We evaluate the same family of methods as in Experiment~2, \rv{adapted to predict at each interaction step rather than only at scenario entry.}
Manual access again serves as the unassisted reference.
Global popularity remains context-agnostic.
Relational retrieval uses the same structured representation, now conditioned on the currently observed context together with the active facet priors at that step.
The zero-shot and few-shot LLM conditions receive the current context and the interaction history and suggest the next facet to present.
For few-shot prompting, we again vary which prior examples are matched to the current step by user, task, environment, or their pairwise combinations.
All methods make 10 suggestions per step.

\subsubsection{\rv{Procedure}}
Following sequence-style evaluation, we compute per-step predictions along each scenario timeline and average at the scenario level so that long scenarios do not dominate the aggregate.
We report both \cost{} and standard top-$10$ retrieval metrics, using the same simulated interaction procedure.

\subsubsection{Results}
\autoref{tab:next-facet-summary} reports the results for next facet suggestion.
As in Experiment~2, the assisted methods reduce \cost{} relative to manual access, but the pattern across baselines differs in informative ways.

Global popularity again provides only a modest improvement (\(-11.7\%\)), suggesting that a generic frequency prior captures only part of the structure needed for sequential support.
Relational retrieval improves more clearly (\(-25.9\%\)), achieving the highest Recall@10, F1@10, and Hit@10.
This indicates that structured inter-facet relations become particularly useful once the task is interactive \rv{and} a richer prior \rv{is available}.
Zero-shot LLM performs \rv{comparably} (\(-24.5\%\)), showing that sequence-aware contextual reasoning can capture some progression structure even without retrieved examples.

The best overall performance comes from few-shot LLM with \emph{User}-matched examples, at the lowest \cost{} (\(-31.1\%\)).
This suggests that next-step functional needs are particularly personalized: once interaction is underway, knowing how a given user tends to proceed appears more informative than the nominal task or environment.
\emph{Environment}-only and \emph{User+Env} examples also perform well, each reducing \cost{} by about \(25\%\), again pointing to situational regularities beyond task labels.

By contrast, \emph{Task}-based transfer is much less effective in this sequential setting.
\emph{Task}-only and \emph{Task+Env} produce only small gains, and \emph{User+Task} performs poorly, with almost no gain over manual access.
This likely reflects data sparsity and over-specific matching in the sequential regime: when few closely matched trajectories \rv{exist}, the examples may \rv{transfer poorly} and even mislead the model.

\paragraph{Qualitative observations}
The same context from Experiment~2 illustrates this difference. 
For (``P04'', ``relaxing'', ``living room''; see~\autoref{fig:qual_reslts} for ground truth). 
Sequential methods, such as relational retrieval, zero-shot LLM, and few-shot LLM with \textit{User} and/or \textit{Env} matches, surface facets missed at initial facet suggestion, including \texttt{Amazon: View Product Details}, \texttt{Search Product}, and \texttt{Open Section}. 
\textit{Task}-based few-shot conditions remain influenced by other users' preferences from similar contexts, and repeated next-step prediction amplifies these mismatches over the scenario.

These results complement Experiment~2 by showing that the most useful source of signal changes once interaction is underway. 
Initial facet suggestion benefits most from matched prior examples, especially user-aligned ones, whereas next facet suggestion can draw much more heavily on evidence revealed by the current trajectory. 
More broadly, \rv{they show how} \system{} supports not only systematic comparison of methods across tasks, but also analysis of how different sources of evidence---global priors, structured history, matched examples, and within-episode interaction---contribute under different adaptive-support settings.

\section{Discussion} \label{sec:discussion}
\rv{In the following, we discuss the implications of our experiments for building context-aware XR systems, and how \system{} can be used and extended.}

\subsection{\rv{Takeaways and Design Recommendations}}
\rv{Our findings suggest several concrete directions for researchers and practitioners designing context-aware XR interfaces.}
\paragraph{\rv{Prioritize user modeling.}} \rv{Across facets, functional needs were user-dominant far more often than task- or environment-dominant (Experiment~1), and user-matched examples gave the strongest gains for initial suggestion (Experiment~2). For deciding \emph{what} to surface, personalization is therefore among the highest-leverage signals, and systems that model individual users are likely to outperform those that adapt only to task or location.}

\paragraph{\rv{Shift the signal source as interaction unfolds.}} 
\rv{The most useful evidence changes over a session: prior user history helps most at cold start, but within-episode interaction becomes the dominant signal once the user begins acting (Experiments~2 and~3). Rather than applying a static context model throughout, a deployed system should lean on user-specific priors at entry and shift toward the live interaction trajectory as it accumulates.}

\paragraph{\rv{Combine context factors rather than relying on one.}} \rv{No single factor accounted for functional needs on its own; user, task, and environment each dominated for different facets (Experiment~1). Systems should fuse these signals rather than key off any one, and the relative weight of each is itself something \system{} can help calibrate on a target dataset.}

% \paragraph{\rv{Match the method to the deployment budget.}}
% \rv{Structured relational retrieval was competitive with LLM-based suggestion while requiring substantially less computation and lower latency (Experiments~2 and~3). For on-device XR, where compute and battery directly affect usability, a lightweight retrieval method can be a better default than an LLM, with generation reserved for cases where its added flexibility justifies the cost.}

\paragraph{\rv{Enrich functional structure at the source.}} 
\rv{Our facets are annotated from existing applications, but richer and more reliable functional structure could come from several routes: app developers declaring functional facets or tags at development time; automated semantic parsing and recognition at the platform or infrastructure level, for those with such access; or generative-interface systems that expose functional units as they synthesize UI~\cite{cao2025generative,chen2025generative,cheng2024biscuit}. 
Investing in this structure, by any of these routes, would make applications more amenable to intelligent, context-aware composition in XR.
}

\subsection{\rv{Using the Framework}}
\rv{\system{} is released as an open-source benchmark with example Python scripts for the three tasks presented in this paper. 
It takes context as input, defined in our benchmark as a triple of user, task, and environment, represented as strings, and evaluates how well a method maps that context to the functionality a user needs.
Context sensing is outside the framework's scope. Practitioners can use their own sensing modules---for example, an activity recognition model or a vision-language model---as long as they produce user, task, and environment labels that then feed into \textsc{MineXR++} and the methods. 
Given this context as input, the framework outputs \cost{} and retrieval metrics for the new method, enabling comparison against the benchmark methods. 
}

% \rv{
% To use the benchmark, a method connects to a single point: it receives a context, and for the sequential task the interaction history so far, and returns a ranked facet set, which the framework scores against the human-authored targets using \cost{} and retrieval metrics across the three tasks. 
% Because the interface simply take context as input and outputs facets, evaluating a method against our baselines requires no changes to the evaluation itself.}

\subsection{\rv{Extending the Framework}} \label{sec:discussion-extending}
\rv{\system{} is designed to be extended along several fronts.}

\paragraph{\rv{Context types.}} \rv{The context representation can grow beyond the user--task--environment triple to include factors such as mobility, social setting, or device type, each of which plausibly shapes what functionality should be surfaced. 
Because methods consume context abstractly, adding a context type requires no change to the evaluation itself.}

\paragraph{\rv{Data.}} \rv{MineXR is derived from lab-based, envisioned layouts rather than long-term in-the-wild use, which limits scale and leaves user- and context-specific signals sparse. 
With larger and more diverse data, \system{} can accommodate techniques that exploit cross-user regularities, such as collaborative filtering and hybrid recommenders, as well as richer ontological representations of users, tasks, and environments. 
Synthetic data from generative agents or context simulators offers a complementary path to expand coverage and stress-test methods on rare or counterfactual scenarios.}

\paragraph{\rv{Tasks.}} \rv{New benchmark tasks, for example longer-horizon prediction or multi-user settings, can be defined over the same facet representation and scored with the same protocol, so that results remain comparable across studies.}

\paragraph{\rv{Metrics.}} \rv{Our \cost{} model captures navigation and search effort under a deliberately simplified simulation. 
This layer can be replaced with richer cost models: biomechanically grounded costs from movement simulation~\cite{belo21xrgonomics,moon2024real,gonzalez2023sensorimotor,miazga2026log2motion}, attention- and cognition-aware costs for the effort of noticing and switching among elements~\cite{lingler2024supporting,Cho2026Sensonaut,cho24auptimize,Li2024Noticeability}, and costs conditioned on a user's familiarity and a device's input and display characteristics~\cite{gajosSUPPLEAutomaticallyGenerating,todi2018familiarisation}. 
Pairing these with qualitative or experience-centered evaluation, including approaches that model user perception and affect~\cite{zhang2024modeling}, would give a fuller account of interface quality than efficiency alone.}

\paragraph{\rv{Methods.}} \rv{Beyond the baselines studied here, \system{} invites richer suggestion methods, for example recommender-style models that exploit cross-user structure or LLM agents that reason over the facet graph, each directly comparable to our results.}

\subsection{Limitations} \label{sec:limitations}
\rv{\system{} emphasizes quantitative, reproducible evaluation, which necessarily abstracts away aspects of experience such as perceived relevance, cognitive load, trust, and interface coherence. 
The \cost{} model is a simplified simulation with fixed per-step costs and a uniform user strategy, and does not capture the full perceptual, cognitive, and physical dynamics of real XR interaction; richer cost models are a natural extension, as discussed above. 
Finally, our evaluation assumes context is given rather than sensed, so it does not account for upstream recognition error; evaluating methods under imperfect, sensed context is an important direction for assessing real-world robustness.}

\section{Conclusion}
We presented \system{}, a benchmarking framework for studying context-aware functional suggestion in XR. 
By introducing functional facets as an intermediate representation of application functionality, augmenting MineXR with facet-level annotations to produce \textsc{MineXR++}, and defining benchmark tasks for \rv{context-factor analysis and} initial and sequential facet suggestion, the framework makes context-aware XR interface support systematically comparable across methods, data, and metrics.
Across three experiments, we showed how \system{} \rv{supports} not only comparison of methods, but also analysis of what drives effective support.
\rv{We found} that desired functionality depends on multiple context factors, and that the usefulness of different signals---global priors, matched prior examples, and within-episode interaction history---varies across adaptive-support settings. 
By grounding evaluation in \cost{}, the framework connects model performance to user-facing effort, moving beyond exact-match metrics toward a more practical notion of utility.
More broadly, \system{} offers a foundation for cumulative progress in context-aware XR systems. 
\rv{By providing a shared representation, tasks, and evaluation, it lets researchers and developers study how different forms of context, data, and modeling approaches help reduce the effort of functional access in XR.}
% By standardizing representation, tasks, and evaluation, it enables researchers and developers to systematically explore how different forms of context, data, and modeling approaches contribute to reducing the friction of functional access in XR.
\begin{acks}
The work in this paper was supported in part by the National
Science Foundation under Award No. IIS-2542186.
We thank Jeongin Park for her help with the figures.
\end{acks}

\bibliographystyle{ACM-Reference-Format}
\bibliography{references}

\appendix
\onecolumn
\section{LLM Suggestion Prompts} \label{sec:prompts}
% For initial facet suggestion with zero-shot LLM, this prompt was used:
\begin{promptbox}
\# Initial facet suggestion with zero-shot LLM \\
\textbf{SYSTEM:} You are an XR (Extended Reality) context-aware functional access support system. \\

You are given a catalog of facet candidates. Each facet represents a user-facing functionality of an app.

A facet is an intent-level functional unit, typically expressed as a verb + noun pair. \\

Examples:

- spotify.view\_now\_playing: checking the currently playing media

- spotify.browse\_music: browsing and exploring music

- slack.browse\_channels: navigating Slack channels

- slack.browse\_messages: reading message content within a channel

Each facet also has capabilities. Capabilities are atomic functional building blocks inside an app, such as APIs, tools, or minimal interaction units.
Some capabilities may have the form open\_<facet\_name>. This indicates that the current facet can directly open or navigate to another facet.
The catalog may expose these navigation edges explicitly through linked\_facets and whether a facet is\_entry\_facet. \\

Facets may also form a facet graph:

- some facets are broader entry or navigation points

- some facets are more specific, deeper functionalities

- "Open Section" is a special entry/root facet representing the app's starting point or navigation hub \\

Use this graph structure to reason about granularity and navigational usefulness:

- prefer the facet whose granularity best matches the user's likely immediate need

- recommend a broad entry facet when the user likely needs general access or exploration within an app

- recommend a deeper facet when the context suggests a specific functionality directly

- recommend adjacent parent/child facets together only if they serve distinct roles in the scenario

- avoid filling the list with redundant facets that are merely consecutive steps in the same navigation chain

- use open\_<facet\_name> capabilities and graph links to infer what is reachable from what

- use linked\_facets and is\_entry\_facet to reason about navigation hubs, nearby functionalities, and non-redundant coverage \\

Your job is to predict a set of 10 functional facets most likely to be useful for the user's current Task and Environment. \\

Consider the following types of usefulness:

- primary: direct support for the immediate task

- secondary: supportive or peripheral actions that help the task

- ambient: low-attention or background utilities the user may keep around \\

Selection guidance:

- prioritize facets that would be useful to surface directly in XR

- balance specificity with breadth across likely needs

- do not over-concentrate on a single app unless the context strongly calls for it

- when multiple facets from the same app are plausible, choose the smallest set that gives meaningful access without unnecessary redundancy\\

Rules:

- Recommend exactly 10 unique facets whenever the catalog contains at least 10.

- Use only canonical facet\_id values from the provided catalog.

- Use the facet graph only to improve relevance and coherence; do not output graph structure.

- If a specific deep facet is recommended, ensure the choice is justified by the task and environment rather than by graph proximity alone.\\

Return JSON only: \{

 "summary": "brief strategy explanation",
 
 "recommendations": [\{
   
     "facet\_id": "canonical.facet\_id",
     
     "tier": "primary|secondary|ambient",
     
     "confidence": 0.0,
     
     "reason": "short explanation"
     
   \}]\}

\end{promptbox}

\begin{promptbox}
\textbf{USER:}

Current scenario:

Task: \{task\_id\}

Environment: \{environment\_id\}

\{details\_block\}

Available facet catalog:

\{catalog\_text\}

Instructions:

- Predict the \{top\_k\} facets most likely to be useful.

- Cover primary, secondary, and ambient support when plausible.

- Use linked\_facets and entry facets when they help explain likely navigation paths or adjacent utilities.

- Use exact facet\_id values from the catalog.

- Return JSON only.

\end{promptbox}

\begin{promptbox}
\# Initial facet suggestion with few-shot LLM \\
\textbf{SYSTEM:} You are an XR (Extended Reality) context-aware functional access support system. \\

You are given:

1. a catalog of facet candidates, where each facet represents a user-facing functionality of an app

2. historical examples selected under a controlled transfer condition \\

A facet is an intent-level functional unit, typically expressed as a verb + noun pair.

Each facet also has capabilities. Capabilities are atomic functional building blocks.

Some capabilities may have the form open\_<facet\_name>, indicating that the current facet can directly open or navigate to another facet.

The catalog may expose these navigation edges explicitly through linked\_facets and whether a facet is\_entry\_facet. \\

Facets may also form a facet graph:

- some facets are broad entry/navigation points

- some facets are deeper, more specific functionalities

- "Open Section" is a special entry/root facet representing the app's starting point or navigation hub \\

Use the graph structure to reason about navigational usefulness and granularity:

- prefer the facet whose granularity best matches the likely immediate need

- use entry/root facets when the user likely needs broad app access or exploration

- use deeper facets when the target context suggests a specific action or information need

- avoid redundant recommendations that are only adjacent steps along the same chain unless they serve distinct roles

- use open\_<facet\_name> capabilities and graph links to infer likely bundles of reachable or complementary functionality

- use linked\_facets and is\_entry\_facet to reason about navigation hubs, nearby functionalities, and non-redundant coverage \\

Your job is to predict a set of 10 functional facets most likely to be useful for the user's current Task and Environment. \\

Consider the following types of usefulness:

- primary: direct support for the immediate task

- secondary: supportive or peripheral actions that help the task

- ambient: low-attention or background utilities the user may keep around \\

Use the historical examples as transfer priors:

- infer stable patterns from the examples

- use them to calibrate likely primary, secondary, and ambient facets

- do not copy examples blindly

- when examples conflict with the current Task and Environment, prioritize the current context

- use the examples to learn likely breadth, specificity, and app/facet combinations \\

Selection guidance:

- prioritize facets that would be useful to surface directly in XR

- balance specificity with breadth across likely needs

- do not over-concentrate on a single app unless strongly supported by both context and examples

- when multiple facets from the same app are plausible, choose the smallest non-redundant set that gives meaningful access \\

Rules:

- Recommend exactly 10 unique facets whenever the catalog contains at least 10.

- Use only canonical facet\_id values from the provided catalog.

- Treat the examples as weak priors, not ground-truth labels for the target case.

- Use graph structure to improve coherence, not to mechanically reproduce chains from examples.

- If a specific deep facet is recommended, ensure it is supported by the target context or transfer evidence, not by graph adjacency alone. \\

Return JSON only: \{

 "summary": "brief strategy explanation",
 
 "recommendations": [\{
   
     "facet\_id": "canonical.facet\_id",
     
     "tier": "primary|secondary|ambient",
     
     "confidence": 0.0,

     "reason": "short explanation"
     
   \}]\}

\end{promptbox}

\begin{promptbox}
    \textbf{USER: }
    
    Current scenario:

Task: \{task\_id\}

Environment: \{environment\_id\}

\{details\_block\}

Transfer condition: \{condition.label\}

Interpretation: \{condition.description\}

Historical examples for this condition:

\{support\_block\}

Available facet catalog:

\{catalog\_text\}

Instructions:

- Predict the \{top\_k\} facets most likely to be useful.

- Cover primary, secondary, and ambient support when plausible.

- Use linked\_facets and entry facets when they help explain likely navigation paths or adjacent utilities.

- Use exact facet\_id values from the catalog.

- Return JSON only. \\

\textbf{SUPPORT BLOCK:} 

Example \{index\}: scenario=\{scenario\_key\}
   
   User: \{participant\_id\}
   
   Task: \{task\_id\}
   
   Environment: \{environment\_id\}
   
   Used facets:
	
    \{app\_name\} / \{facet\_name\}
	
    [\{facet\_id\}]

\end{promptbox}

For next-facet suggestion, the same prompts as initial facet suggestion were used, but the following was added to the context:

\begin{promptbox}
Existing user-placed facets in the XR layout:

[

\{idx\}. facet\_id=\{facet\_id\}; app=\{app\_name\};

facet\_name=\{facet\_name\}; relevance=\{rel\}

]

Use these already placed facets as prior context, avoid repeating them, 
and recommend what should come next for the remaining needs.
\end{promptbox}

\end{document}